# Software-Defined Network-Based Vehicular Networks: A Position Paper on Their Modeling and Implementation

**Lionel Nkenyereye [1], Lewis Nkenyereye [2,\*], S. M. Riazul Islam [3], Yoon-Ho Choi [4], Muhammad Bilal [5]** 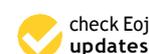 **and Jong-Wook Jang [1,\*]**

1. Department of Computer Engineering, Dong-Eui University, Busan 614-714, Korea
2. Department of Computer and Information Security, Sejong University, Seoul 05006, Korea
3. Department of Computer Science and Engineering, Sejong University, Seoul 05006, Korea
4. Division of Computer and Electronics Systems Engineering, Hankuk University of Foreign Studies, Yongin-si 17035, Korea
5. School of Computer Science and Engineering, Pusan National University, Busan 46241, Korea
* Correspondence: nkenyele@sejong.ac.kr (L.N.); jwjang@deu.ac.kr (J.-W.J.)

**Abstract:** There is a strong devotion in the automotive industry to be part of a wider progression towards the Fifth Generation (5G) era. In-vehicle integration costs between cellular and vehicle-to-vehicle networks using Dedicated Short Range Communication could be avoided by adopting Cellular Vehicle-to-Everything (C-V2X) technology with the possibility to re-use the existing mobile network infrastructure. More and more, with the emergence of Software Defined Networks, the flexibility and the programmability of the network have not only impacted the design of new vehicular network architectures but also the implementation of V2X services in future intelligent transportation systems. In this paper, we define the concepts that help evaluate software-defined-based vehicular network systems in the literature based on their modeling and implementation schemes. We first overview the current studies available in the literature on C-V2X technology in support of V2X applications. We then present the different architectures and their underlying system models for LTE-V2X communications. We later describe the key ideas of software-defined networks and their concepts for V2X services. Lastly, we provide a comparative analysis of existing SDN-based vehicular network system grouped according to their modeling and simulation concepts. We provide a discussion and highlight vehicular ad-hoc networks' challenges handled by SDN-based vehicular networks.

## 1. Introduction

Vehicle-to-everything (V2X) communications are definite technologies in vehicular networks to drastically reduce road accidents and enable a high-level of vehicle automation. For years, the technology of choice for V2X, on one hand, has been Dedicated Short Range Communication (DSRC) [1], which is based on IEEE802.11p technology [1,2]. On the other hand, Cellular-V2X (C-V2X) technology is seen as a new communication standard supporting V2X services [3]. LTE-V2X technology is a derivative of the cellular uplink technology that maintains similarity with the current LTE systems [2]. Furthermore, the focus on V2X technology expands the availability of a wide range of services that include cloud-based vehicular services and edge computing [3]. Therefore, vehicles access these cloud-based services through road side units (RSUs). Thus, RSUs increase the reliability of disseminating critical safety messages to a large number of vehicles [4].

RSUs are communication nodes with the vehicular networks. This means that the vehicle needs to have access to road infrastructures through RSUs using infrastructure-based communications (hereafter V2I) [5]. For instance, RSUs forward received messages to intelligent transportations system (ITS) application servers by exploiting wide area networks [5]. Although communication capabilities between vehicles depend highly on the number of RSUs deployed and their coverage, RSUs are surely costly to deploy and to maintain. Consequently; there is a trade-off between full connectivity through RSUs and the deployment cost. To overcome the deployment cost of RSUs, road operators (ROs) can additionally leverage spectrum owned by mobile network operators (MNOs) to control traffic management services. In this situation, ROs are certainly expected to deploy and manage public-sector RSUs [6]. Following this, the ROs can enter into business arrangements with MNOs to surely deploy RSUs and run V2X services provided by ITS's authorities [6]. Therefore, MNOs should leverage existing cellular infrastructure to promote efficient deployment of V2X services.

Though the IEEE 802.11p was tested, automotive makers have manifested interest in C-V2X technology and question the applicability of the IEEE 802.11p for enabling many new V2X services. These doubts about the use of IEEE 802.11p coincides with the emerging of the fifth generation (5G) technology which aims to reduce network management through automation [7]. Furthermore, the commitment of automotive OEMs to test cellular communication for V2X motivated them to be part of a wider progression of 5G era [7]. The key technology of 5G design is mainly focused on the automation of network resources by using network slicing [8] which in turn is based on two new network technologies: network function virtualization (NFV) and software-defined networks (SDNs) [9]. The SDN concept together with edge computing could resolve most issues in vehicular networks such as irregular connectivity packet loss rate [8,10]. Therefore, software-defined-based vehicular network (SDVN) systems [8,10] improve resource utilization, selection of best routes, and facilitate network programming [9]. These SDVN architectures define local SDN domains through clustering in order to access the global intelligence of the network managed by the SDN controller [11,12].

There is a considerable amount of research work on SDVN [8–12] that focuses on different concepts, including the definition of SDN, software entities of the control plane, routing protocols using SDN-based VANET, etc. Some authors have proposed innovative architectures based on existing V2X scenarios that provide optimization results of their proposed architecture. There is also a number of surveys [13,14] that summarize the current work in the literature. However, it is quite challenging for most of the researchers to quickly decide which proposed solution could be suitable for their use case from schemes that propose modeling, architecture, and optimization.

In this paper, we provide a review of published articles in the literature to comprehend the present state of research concerning software-defined networks-based vehicular networks with a particular focus on the articles whose contributions include modeling and implementation. Consequently, we performed a search on Google Scholar with the following keywords: software-defined networks, software-defined networks-based vehicular networks and modeling and implementation. In addition, we used the same keywords on other three research web engines, namely ScienceDirect, IEEE and ACM. Since SDN and VANETs are relatively new topics, we did not retrieve a huge number of papers that required an established protocol for evaluation and selection. Therefore, articles were manually selected or excluded if a given article provides clear modeling and implementation techniques. Other criteria were used in the selection such as significance, citation or rank of the publication venue.

In this work, we mainly focus on providing implicit literature that focuses on classifying existing SDVN solutions based on their modeling and implementation. To the best of our knowledge, it is the first work that groups SDVNs based on their modeling and implementation schemes. Therefore, in this paper the main contributions are summarized as follows:

- We first overview the current studies available in the literature on C-V2X technology in support of V2X applications.

- We then present the different architectures and their underlying system model for LTE-V2X communications.
- We also describe the keys ideas of software-defined networks and their concepts for V2X services.
- We define four elements that are considered for modeling and implementations of SDN for vehicular networks. We then present a comparative analysis for existing schemes grouped according to their modeling and simulation concepts.
- We provide a discussion and highlight vehicular adhoc network(VANET)'s challenges handled by SDN based vehicular network.

The remainder of the paper is organized as follows: the current studies and technologies for V2X services are detailed in Section 2. A comparative study of architectures and a system model of LTE-V2X communication in the implementation of V2X services are discussed in Section 3. The modeling and implementation of software-defined vehicular networks for V2X is detailed in Section 4, together with a definition of SDN, before briefly discussing findings on the comparative study of existing SDN based vehicular network in Section 5. Finally we conclude our work in Section 6.

## 2. Current Studies and Technologies for V2X Services

This section relates the evolution of vehicles equipped either with IEEE 802.11 p or C-V2X wireless communication technologies for deploying V2X services. This section describes the V2X and C-V2X communications modes. A comparative study of existing architectures and a system model of LTE-V2X communication in the implementation of V2X services are detailed.

*2.1. V2X Communication Modes*

A vehicle can interact with its environment through various types of communication as specified in [15]:

(1) Vehicle-to-Vehicle (V2V): A type of communication, in which User Equipements (UEs) (such as vehicles) communicate using V2V services.
(2) Vehicle-to-Pedestrian (V2P): A type of communication, in which both UEs (vehicle, pedestrian) communicate using V2P services.
(3) Vehicle-to-Infrastructure (V2I): A type of communication, in which one part is a vehicle- capable user equipement (VUE) and an RSU entity, both communicating using V2I services.
(4) Vehicle-to-Network (V2N): A type of communication, in which one part is vehicle-capable user equipment (VUE) and the other part is a V2X application server on the cloud for instance, both communicating using V2N services. As shown in Figure 1, V2N relates to any communication between vehicles and computing infrastructures such as RSU deployed either with eNodeB or like a standalone stationary UE [15].

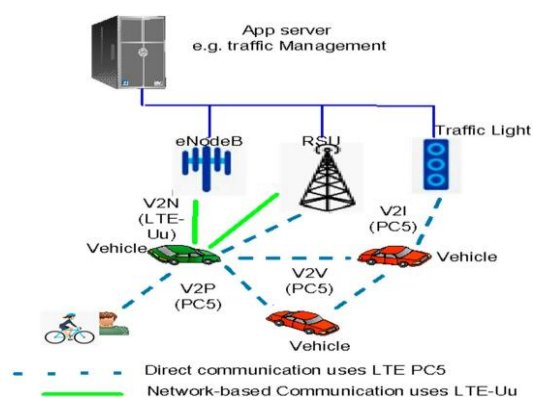

**Figure 1.** 3GPP Release 14 [16] for V2X services using direct communication over side link PC5 and LTE-Uu.

*2.2. Evolution of Vehicles Using V2X Services*

The study on the socio-economic benefits of cellular V2X [17] conducted by "The Analysys Mason" [17] specifies four (4) case scenarios to study the evolution of vehicles either equipped with IEEE802.11 or C-V2X technologies for deploying V2X services. These case scenarios are numbered from one (1) to four (4). Scenario one (1) is the case adoption of C-V2X and IEEE 802.11p in the absence of any government measures. The second scenario is the case all new vehicles to support ITS services using IEEE 802.11 p in 2020; the third scenario is the case in 2023 all new vehicles are equipped with LTE PC5. The fourth scenario is the case the Equitable 5.9GHz use is adopted for V2X communications.

Lessons learned from the study in [17] are described in Table 1, which summarizes case scenarios about V2X communications and relevant challenges. In the absence of any government regulations, V2V would use IEEE802.11p or LTE-V2X PC5 [18]. This means that no-direct communication interoperability between IEEE 802.11p and PC5 exists. Therefore, V2V is possible via cellular LTE and vehicles without IEEE802.11p or PC5 will use V2I and V2P via LTE-Uu of a smartphone brought in the vehicle. The case scenario 2 concerns all new vehicles that will use IEEE 802.11p in 2020 to support ITS services. Although vehicles without IEEE 802.11p would not communicate via V2V and V2I, vehicles equipped with IEEE 802.11p and LTE Uu could communicate via the cellular network. The challenge of the scenario case 3 (all new vehicles equipped with LTE PC5) would dictate ROs to add PC5-based RSU to existing RSUs potential. This means that vehicles without PC5 enabled would have to use V2I via smartphone. The case scenario 4 that predicts the use of Equitable 5.9GHz would allow automotive OEMs to use IEEE 802.11 p for V2V/V2I and Cellular (LTE-Uu) for V2N. In conclusion, the base case (case scenario 1) and equitable 5.9GHz RSU (case scenario 4) [19] deployment are thus suggested to be the most profitable way to deploy V2X services based on the net benefit perspective.

**Table 1.** Use Case scenarios to study the penetration of V2X services. The study was carried out by Analysys Mason [17].

| Scenario# | Description | Vehicular Communication | Remarks |
| --- | --- | --- | --- |
| Base case | Adoption of C-V2X and IEEE 802.11p in the absence of any government measures | V2V using IEEE802.11p or LTE-V2X PC5 | V2V is possible via cellular LTE and V2I and V2P via LTE-Uu of a smartphone |
| Scenario 2 | In 2020, all new vehicles to support ITS services via IEEE 2020 | IEEE 802.11p for V2V and V2I | Road operators should install new RSUs or expand them to support V2I |
| Scenario 3 | In 2023, all new vehicles equipped with LTE PC5 | V2V and V2I via LTE PC5 | Road operators add PC5-based RSU to existing RSUs |
| Scenario 4 | Equitable 5.9GHz use | Division spectrum for V2V based PC5 and IEEE 802.11p | IEEE 802.11 p for V2V/V2I, Cellular(LTE-Uu) for V2N and others use PC5 for V2V/V2I |

The number of vehicles equipped with embedded C-V2X communication technology is expected to increase as shown in Figure 2. Even though after a while we would expect C-V2X to be equipped in a greater number of vehicles, vehicles which do not have embedded C-V2X would use LTE PC5-based smartphones for accessing V2I and V2P services. In this context, the base case (scenario 1) seems to be the one to be adopted by many automotive makers. A key challenge in this scenario is predicted when different automotive OEMs would deploy different V2V communication solutions. Consequently; the inefficient use of the equitable 5.9GHz spectrum could occur due to no direct-communication interoperability of the two technologies (C-V2X and IEEE802.11p).

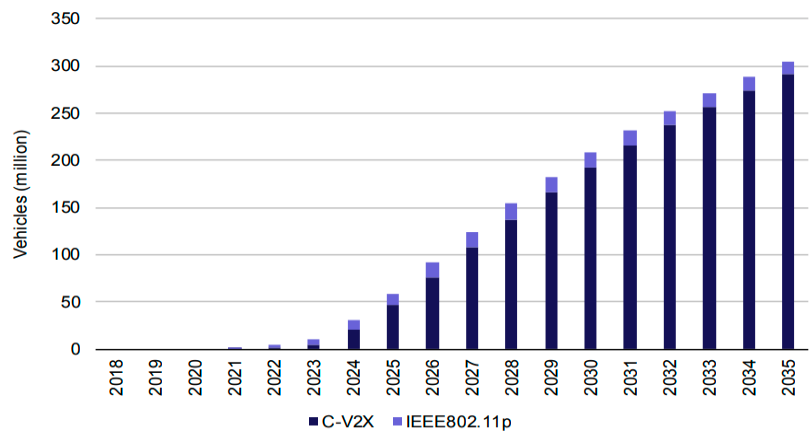

**Figure 2.** Evolution of the number of vehicles using V2X services in base case scenario 1 (Table 1) of vehicular technology [17].

*2.3. 3rd Generation Partnership Project (3GPP) Cellular-V2X*

3GPP Release 14 [16] for V2X services using direct communication over side link PC5 and LTE-Uu is shown in Figure 2. Direct communication uses links over the side link PC5 reference interface. In fact, side link PC5 defines features based on proximity service (ProSe) which is adapted for V2V communication scenarios [16]. PC5 communication mode enables V2I communication between vehicles and road infrastructures such as traffic control lights. In addition, V2N service uses LTE-Uu for allowing communication between vehicle and computing infrastructures, for example, an RSU implemented either with an eNodeB or as a standalone stationary UE, central cloud computing. A vehicular enabled UE exchanges data with deployed computing infrastructures over the LTE-Uu interface through RSU. The RSU broadcasts V2X messages towards multiple vehicles enabled UEs in a target area through the evolved multimedia broadcast multicast service (eMBMS) [16]. V2N serves VUEs in communication with an application server hosting ITS management applications, referred as a V2X Application Server (AS), which would provide a global state of traffic, the management of it, and service information [16,19–22].

Actually, cellular communication today represents the most embraced solution to collect data from vehicles and retransmit them to the network through RSUs. This avoids having to build new or set-ups expensive installations of RSUs [23,24]. To address admittedly V2X services use cases, the technical specification group (TSG), radio access networked (RAN) define V2V service using device-to-device (D2D) as specified in Release 12 [21]. Thus, a direct communication interface called sidelink (or PC5 interface) was thereafter specified in Release 14 [16] to allow direct communication link between devices. In addition, improvements to this interface have been added within Release 14 to study the V2V use cases in the ITS 5.9 GHz band and more specifically in [22].

**3. A Comparative Study of Architectures and a System Model of LTE-V2X Communication in the Implementation of V2X Services**

Important research on LTE-V2X communication in implementing V2X services has started to show relevant results. The relevant results of existing works focus mostly on the following network concepts: (i) long-term evolution-vehicle (LTE-V) standard supporting V2V communications using PC5 in LTE [25], (ii) methodical and assimilated V2X solution based on time-division LTE (TD-LTE) [26], (iii) multi-channel licensed-assisted access (LAA) schemes to enlarge multi-carrier Wi-Fi network [27]. We identify the following categories of work addressing system model of LTE-V2X communication in the implementation of V2X services:

(1) Relevant use cases and requirements for V2X services
(2) Design choices determining the performance of LTE-V2X communications

*3.1. Relevant Use cases and Requirements for V2X Services*

Boban et al. [28] describe the benefits of vehicles cooperating through V2X communication. They define descriptions and requirements of some relevant use cases which would be supported by future V2X communications systems. Among relevant use cases presented, some of them are bandwidth-demanding applications with high link reliability estimated to reach 99%. Considering latency, the authors mentioned that a low latency with a value below 10 ms is required for most of relevant uses cases. Therefore, these relevant uses cases require a high throughput of tens of Mb/s per vehicle. In addition, Seo et al. [4] provide a survey of the service flow and conditions of the V2X services based on LTE systems. They also discuss relevant scenarios suitable for an operational LTE-based V2X services system. Their work reveals some challenges such as high mobility and high density of vehicle which would bring a great impact in designing practical and technical solutions to satisfy the requirements of V2X services.

*3.2. Design Choices Determining the Performance of LTE-V2X Communication*

Masegosa and al. [25] put forward an overview of the long-term evolution-vehicle (LTE-V) standard supporting V2V communications using LTE's direct interface known as PC5 in LTE. The overview of physical layers changes presented under release 14 for LTE-V allows both communications modes 3 and 4 of the LTE-V. LTE-V is under study and its specifications would be published in Release 15 [6]. This Release 15 defines specifications on fifth-generation (5G) for supporting both V2X services and self-driving vehicles' applications. Indeed, the goal of Masegosa and al.'s work [25] was to review V2X Communications under mode 3 and mode 4 with LTE-V. In mode 3, the resources are assigned by the cellular network while mode 4 does not depend on cellular coverage, and vehicles autonomously take their radio resources using a relegated scheduling scheme supported by congestion control.

The results of the works in [25] discusses the performance achieved by the most major wireless technology IEEE 802.11p compared to LTE-V when vehicles transmit 10 packets per second (pps) to a distance of 160 m. In case the 802.11p data rate is increased to 18 Mb/s to a distance up to 160m, IEEE 802.11p achieves a smaller packets data rate (PDR) than LTE-V thanks to the physical layer performance and the overriding effect of propagation. The authors analyzed also the performance of (LTE-V) standard when the channel load increases, this means when a vehicle transmits 50 packets per second (pps); the results show that the packet collisions become the primary source of errors.

Chen et al., [26] put forward a long-term evolution (LTE)-V model with a contribution on a methodical and assimilated V2X solution based on time-division LTE (TD-LTE). The main idea is the use of a centralized architecture that highlights features of TD-LTE and LTE-V-cell optimizes radio resource management for supporting better V2I. The results from their study are compared with the well-known wireless technology, IEEE 802.11p. The comparison reveals that LTE-V inherits the advantages of TD-LTE, including local features of TD-LTE and LTE-V-cell for supporting V2I communication implemented based on a centralized architecture. Therefore, they suggested that LTE-V would consort new features to overcome the challenges of V2V communications, such as congestion control.

Mukherjee et al. [27] studied the impact of unlicensed spectrum operation on the LTE physical layer architecture and the study of farther enhancements about licensed-assisted access (LAA). They present a brief survey of valuables enhancements for LAA for upcoming LTE releases. The experimental results of their proposed system expose clearly that from the synchronization point of analysis and the influence on the non-substitute Wi-Fi network, both classes of multi-channel LAA LBT schemes are realizable and can enlarge the performance of a multi-carrier Wi-Fi network assimilated when it is synchronizing with another Wi-Fi network.

Kawasaki et al. [29] proposed a performance evaluation between two methods of LTE-based V2X. The two methods are Uu-based LTE-V2X based and PC5-based LTE-V2X which is supported by device to device (D2D) communication [22]. The authors argue that queuing latency is significantly affected by bandwidth allocation, latency, parallel degree (PD) both in PC5-based and Uu-based. The authors reveal that the numbers of admissible parallel transfer are decided by different factors in Uu-based and PC5-based LTE V2X. However, in case the number of parallel transfer is equivalent to a larger logical bandwidth, queuing latency is estimated to remain smaller. The experimental evaluation results show that at PD=8, Uu-based was recorded to have the latency of 69.91msec and PC5-based LTE to have a latency of 11.82msec. To sum up, the latency of PC5-based had only 16.9% of the latency in Uu-based. PC5-based LTE unveiled to retain a better performance than Uu-based while PC5-based requires additional functions compared to the existing LTE.

## 4. Modeling and Implementations of Software-Defined Vehicular Networks for V2X

*4.1. Definition of Software-Defined Networks*

Software-defined networks (SDNs) [30] are based on the separation of data and control planes. In SDNs, communication between the control layer and network layer takes place through the SDN control protocol because the control plane and forwarding plane are decoupled. Based on this principle of decoupling data and control plane, a standard protocol with multivendor support was needed for enabling communication between SDN's layers. As a result, OpenFlow was developed for this purpose [30]. OpenFlow was the first open-source control protocol for communicating between the SDN controller and the network devices. OpenFlow enables the implementation of a user application program to manipulate directly network devices without implementing various network protocols. Furthermore, OpenFlow maintains what it calls a flow table [31] on the network device (forwarding devices). The flow table contains information on how the data needs to be forwarded [30]. The SDN controller can then use OpenFlow to program the network devices of an OpenFlow-enabled switch by altering this flow table [32]. To program the forwarding information and set up the path across the network, the OpenFlow architecture supports two modes of operation, reactive and proactive [33]. The reactive mode is the default method of implementing SDN using OpenFlow and assumes that there is no intelligence of a control layer running on the network devices. In this mode, the first packet of the data traffic received on any of the forwarding nodes is sent to the SDN controller, and then the SDN controller uses this information to program the flow across of the whole network. In proactive mode, the SDN controller is preconfigured with some default flow values, and the traffic flow is programmed preemptively as soon as the switch is brought up. SDN controller and switches exchange the flow of information over the network using a secure channel such as Secure Socket Layer (SSL) or Transport Layer Security (TLS) while the OpenFlow manages communication between network layer and control layers [34,35].

*4.2. Software-Defined Networks and their Concept in Vehicular Networks for Deploying V2X Services*

The control layer plane is responsible for collecting and maintaining the status of all SDN cellular network devices, RSUs, and the vehicles [8]. An example of such SDN deployment in V2X services could be the route prediction on demand. The application could monitor vehicles on the roads and provides additional route prediction paths at a certain time of the day or when the vehicles are temporarily disconnected due to the high speed of the vehicles. The control layer would have to provide with the information about the vehicle's future route based on the Global Positioning System (GPS) or a navigation system [11]. The ability to deploy V2X services through SDN concepts is perhaps the most significant for automakers to solve the challenges of the no-direct interoperability of vehicle's wireless interface. Today, deployment of V2X services demands higher agility in network restoration, massive scalability, faster deployment, and operating expense optimization [36]. Therefore, V2X services cannot simply afford to be slowed down by the lack of speed in human-driven processes.

Automakers' onboard wireless communication interfaces have been traditionally specific to their vehicles. Automakers offer limited support for allowing external network devices to make decisions based on the logic and constraints across the vehicular networks. SDN offers a solution by linking V2X services to the vehicular network and bridging the challenge that existed with manual control and management processes. In addition, maximum use of automated tools and application have become a necessity to meet the V2X service demands. Automation and programmability capability are needed to support the provisioning of V2X services, the monitoring, and interpreting of V2X networks devices data. Therefore, automated tools implement run-time changes based on high mobility of vehicular networks, road traffic loads, and disconnection due to a high speed.

Since the SDN puts the intelligence of the vehicular networks in a central controlling software called SDN controller which conveys vehicular routing protocols to VANET's wireless nodes (such vehicles, RSUs). In fact, the vehicular routing protocols automatically react to the vehicle's mobility since the global view of the network is permanently available on SDN Controller. Therefore, the dissemination of routing path based on the vehicle's speed could be built directly into the SDN controller. Alternatively, the open protocols to manage the V2X applications can run on the top of the SDN controller using the northbound bound APIS [30] to proceed down the routing policies and rules to the controller and southbound APIS [30] to convey routing policies from SDN controller to the V2X forwarding devices. In conclusion, features of the SDN should handle the issue of high mobility and then improve V2X messages exchanged in a heterogeneous VANET architecture.

### 4.3. Architecture Overview of Software-Defined Vehicular Networks

Figure 3 depicts the components of various wireless communications in the software-defined vehicular network. To allow an SDN-based vehicular network (SDVN), simulation conducted on it leads to a certain number of SDN components. The SDN controller is the central logical intelligence of the SDN-based vehicular system. The SDN controller has a generalized and global view of the vehicular network and implements Openflow protocol to handle routing policies to eNB-type RSU controller on RSU. In fact, eNB-type RSU controller deployed on the edge of the vehicular network shortens the decision of generating new routing packets undefined in forwarding devices' flow tables. The SDN controller V2X network management conveys routing policies to UEs (vehicles) by implementing ITS's goals set up on the cloud or at the edge of the network for lowering processing decisions. The SDN controller is not only responsible to provide the whole performance but also provide routing rules for wireless devices (vehicles) selecting best routing paths to their destinations in VANET. OpenFlow enabled V2X-EU is the SDN wireless node and is responsible to control the data plane elements [12]. Data plane on vehicle implements OpenFlow protocol and is embedded in the OnBoard Diagnostic Unit (OBU). Furthermore, data plane elements are the VUEs that perform control message in term of routing policies from the eNB-type RSU controller to execute predefined actions which state ITS's goals once implemented in the application plane of the SDVN.

### 4.4. Modeling and Implementations of SDN for Vehicular Networks

The study of existing works on SDN-based vehicular networks was conducted based on the four (4) basics elements of modeling and simulation scheme [37]. First, we identified the targeted drawback that the researchers addressed. The second element is the classification of the existing SDVNs or VANETs system on which belong the addressed drawback. The third is the systems analysis which allows identifying parts of the SDVN system that are relevant to the problem. Finally, the model of the proposed solution, in turn, provides the implementation of the model related to the SDVN system in considering the outputs of its system analysis. Modeling and simulation scheme of existing work on SDVN was proposed to study the issues of several problems that originate from the complexity of ITS's applications understudy in VANETs and Internet of Vehicles. Thus, the models of software-defined vehicular networks contribute as a network technology to provide a solution to current VANETs' applications. In addition, SDVN is considered as a system because it is a part of VANET technology that will influence the design of future vehicular network architectures. The summary of the modeling and simulation schemes of existing works on SDN-based vehicular networks is described in Table 2.

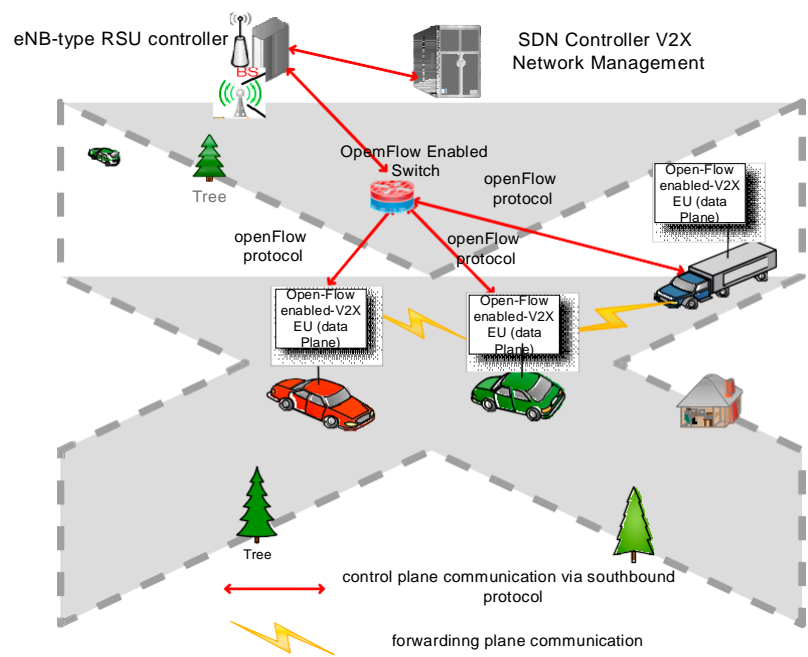

**Figure 3.** Software-Defined Vehicular Network. Data plane on vehicle implements OpenFlow protocol and is embedded in the OnBoard Diagnostic Unit (OBU). The SDN controller has a Generalized Vehicular cloud Openflow Controller on RSU. The SDN controller conveys routing policies to UEs (Vehicles) by implementing ITS's goals set up on the cloud.

Mainly, a VANET deals with systems in its objectives in a way of filling the separation between heterogeneity caused by communications interfaces equipped in vehicles or infrastructure-based communication. For instance, let us consider the fact that high mobility of vehicles causes dynamic topology change that in turn generates packet losses in the network, therefore routing protocols in mobile entities to effectively handle the short lifetime link are required. To this, the modeling and implementation of this issue conducted by the authors in [12] and summarized in Table 2 show that it is an SDVN's challenge and the research community suggests what could be done to solve the problem. The system analysis which represents the entities of the system that are relevant to the problem discloses trace of message overhead between vehicles (data planes entities) and the SDN controller. To this end, researchers should quickly decode that message overhead between vehicles (data plane) and SDN controller is the root cause, therefore, the implementation of the solution to a new problem related to routing protocols that could break link quality would start on message overhead on the SDN controller. Thus, the model proposed by the authors in [12] involves a new routing protocol that improves the packet delivery ratio by selecting stable routes with the lowest latency to control the overhead message on the SDN controller. Inalterability in protocol deployment due to the heterogeneity of wireless infrastructures prompted the authors in [38] to provide a system analysis that centers on abstracting heterogeneous wireless nodes as SDN switches enabled OpenFlow and designing SDN controller to manage dynamically network resources.

The output of the system analysis prompts the authors in [38] to propose a solution model that includes an adaptive protocol for heterogeneous multihop routing, a topology that enables SDN management overhead via the status of SDN switches and finally provide use cases of SDVN-enabled V2V, V2I, and V2N. To improve the performance in communication by mitigating the connectivity loss between vehicles and central SDN controller in [39], the authors suggested a system analysis based on selecting local SDN controller domains through clustering concepts. They proposed a hierarchical SDNV as the implementation model to decrease connectivity loss at the SDN controller, consequently; enhance the robustness of internetworking of data plane entities.

**Table 2.** Summary of related works on SDN based vehicular networks grouped according to the modeling and implementation scheme.

| Description of the Problem | System | System Analysis | Model of the Proposed Architecture |
|---|---|---|---|
| Connectivity loss between vehicles and SDN controller [39] | SDVN | Local SDN controller domains through clustering | Hierarchical placement of SDN controllers decrease connectivity latency between them |
| Routing in mobile cloud [12] | SDN-based routing | Track message overhead between vehicles and controller | Control the overhead of the SDN controller and packet delivery ratio |
| Amount of data transfered for multimedia applications [1] | SDVN | Analyze throughput, end-to-end delay | RSU micro-datacenter, stochastic switching for reconfiguration overhead |
| Heterogeneity of wireless infrastructures and inalterable in protocol [38] | SDVANETs | Abstract heterogeneous wireless nodes as SDN switches enabled OpenFlow Allocate network resources through SDN controler | Deploy adaptive protocol for heterogeneous multihop routing; mitigate SDN management overhead via status of SDN switches; SDN enabled V2V, V2I and V2N. |
| Efficient resource utilization [11] | Software-defined Cloud/Fog network | SDN supports hybrid mode, Control plane is distributed between SDN controller, BS and RSU | Fog computing concept is adding to provide FSDN |
| Latency control [10] | Software-defined Mobile Edge computing | Software-defined cloud/edge vehicular networking | Latency control mechanisms: radio access steering at the base stations (BSs) |
| Latency control [40] on Multiple core network for autonomous driving vehicle | Software-defined VANET with 5G | Local knowledge of surroundings nodes, SDN controller, Broadcast beacon message | Cellular network integrated with network Model, SDN control eNB infrastructure, RSU controller controls RSU |
| Latency control and cost on cellular network [32] | Software-defined VANET with 5G | Control communication: VANET based, cellular network-based, hybrid-based | Optimize southbound communication via rebating mechanism, game equilibrium, two-stage leader-follower game for best decision between vehicle and controller |
| Dynamic resource management [14] | Software-Defined VANETs | Topology of SDN controller, Model of Node in Mininet-WiFi | Extend modeling of node car in mininet-WiFi |
| Control latency communication [13] | Vehicular networking; heterogeneity of radio access technologies | Vehicle network architecture for resource management, SDN controller, redesign of existing vehicular networks | Model SDHVNet architecture |

ITS scenarios in future VANETS require quality of service (QoS)s and efficient utilization of network resources for enabling autonomous driving. The authors in [11], [14] addressed the challenge of efficient resource utilization. In [11], the system on which the problem is associated use the fog(edge) computing technology, thus the SDN-based fog network is evaluated to propose location-aware services with less communication latency. To this end, the system analysis centers on the deployment of the SDN to support hybrid mode (central-based and distributed-based configuration of SDN controller) and on the configuration of control plane (SDN controller) in distributed mode with both the base station (BS) and RSU. Considering the outputs of the SDN-based fog computing, the authors in [11] propose a model that combines edge(fog) computing services for allowing heterogeneous communication access for V2V, V2I, and V2N. The authors in [14] provide a system analysis that centers of the topology deployment of SDN controller and the possibility to model communication nodes (vehicles) as an SDN switch using the open-source simulation tool known as Mininet-WiFi [14]. The proposed model offers efficient utilization of network resource after modeling the vehicle as a node using Mininet-WiFi. Taking mobile edge computing step further, the authors in [10] investigate the possibility of deploying VANET's application with low-latency and high-reliability communication delay in software-defined mobile edge computing. The system analysis provided by the authors in [10] takes into consideration the edge vehicular network architecture which in turns provide a modeling solution on how to control the communication latency through radio access steering at the base station.

The advancement of 5G in the automotive field brings the integration of VANETS and 5G technology to construct 5G software-defined vehicular network with SDN technology as a primary key enabler. The authors in [32,40] investigated the challenge of communication latency and the cost on multiple core network for the autonomous driving vehicle. The modeling and implementation of [32] provide a systematic analysis based on the control of latency at VANET position, the cellular network- or hybrid-based (VANET and cellular position). The outputs of the system analysis prompt the authors in [32] to model their solution for decreasing communication latency by optimizing southbound communication via both rebating mechanism and the use of game equilibrium associated with the two-stage leader-follower game in order to select best routing paths between vehicle and controller. In [40], the modeling concepts centers on the system analysis based on broadcasting V2V beaconing messages so that the local knowledge of surroundings nodes and their topology are available at the SDN controller. After system analysis, the proposed solution provides a model that includes the integration of SDN controller, eNB infrastructures, RSU controller and 5G to design 5G based SDN concept.

The full transformation of VANET into SDVN requires to model SDVN solutions not only based on system architectures of SDVN but also based on mathematical analysis. Since the SDVN integrates the use of the SDN concept on the VANETs, a mathematically-based model is the natural modeling language to break up complexity problems and make VANETs' and SDVNs' challenges tractable. Mathematical-based theory applied to SDVN should bring further improvements and variations for allowing SDN to fully enhance VANETs, consequently, minimizing latency and cost, safety message delivery using heterogeneous communication interfaces [41]. A thorough mathematical model theory for all the above–analyzed articles that would lead to a new proposed concept that along with its implementation shall be addressed in future work.

## 5. Discussion

In this section, we summarize our findings from the classification of SDVNs based on the modeling and implementation schemes. The modeling strategy used in this paper to break up SDVNs' architecture helps to sort out existing VANETs' challenges addressed by integrating the SDN concept in VANETs. Some of the problems and system analysis in the process of modeling for problem-solving have been covered in Section 4.4, however, the rest of this section covers the summary of the four elements on which we centered the modeling of existing SDVN architecture in order to comprehend the current VANET's challenges solved by SDVN system. The simplicity of modeling proposed in this paper aims at encouraging a research combination towards SDVN with 5G and with edge computing as an alternative solution for future VANET's applications. Based on our study, we provide SDVNs' systems analysis of existing SDVN architecture.

The comparative study of existing SDVN based on the modeling and implementation scheme as shown in Table 2 provides a list of a number of VANET issues addressing the full transformation of VANETs to SDVN. To this end, the identified VANET issues handled over to SDVN systems are summarized in the following contributions: firstly, the contributions in [10,13,32,39,40] that address the issue of loss of connectivity by controlling the data plane latency, secondly, the contributions of authors in [12] that are related to routing protocols in the mobile cloud environment, thirdly, the contributions from authors in [11,14] which address the issue of resource utilization. Finally, the authors in [1] provide a study on the amount of data transferred for multimedia applications. Lastly, the contributions in [32,40] address the issue of communication latency and the cost of using multiple core network for autonomous vehicles.

Moreover, handover control and proper allocation of radio resource were analyzed in [42] to mitigate the challenge of mobility management and transmission delay. The mobility management in VANETs increases delays in the transmission where handover procedures are not properly implemented.

To this, SDVN with fog computing would allow meeting the requirements of low transmission delay by adopting a hybrid handover scheme, optimizing radio resource allocation through the Markov decision process [42]. However, inefficient control for high mobility that causes unsteady wireless channel for SDVN and latency on the distribution of commands from controllers and interworking breach through heterogeneous networks were among ongoing VANETs' challenges to contend with SDVN. In addition, network slicing and NFV [25] in SDVN introduce potential research opportunities. In fact, SDN allows operative network slicing in a dynamic topology. The NFV with the use of hypervisor has the task of adjusting OpenFlow in the way to enable heterogeneous network interworking.

The system analysis of SDVN systems identifies components, architectures, protocols directly linked to the SDVN challenge addressed. Note that there are six (6) architecture systems of SDN-based vehicular network proposed in the literature: SDVN [1,39], SDN-based routing [12], software-defined VANETs (SDVANETs) [14,38], SDN-based cloud/mobile (fog/edge) computing [10,11], software-defined VANET with 5G [32,40]. A comprehensive study of SDVN architecture, its benefits and services are described in [36]. Although six systems of SDVN architectures are currently implemented and simulated, system analysis provides insights to relevant components, architectures, protocols and simulation tools to be considered before providing a solution model to VANET's challenge. In fact, we can list a few of SDVN system's analysis as summarized in Table 2: placement of SDN controller [11,13,14,39,40], communication control VANET-based or cellular network-based [32], local knowledge of surrounding nodes via beacon or geo-broadcast messages [39,42], network simulator tools such as Mininet-WiFi [14], trace of overhead messages between vehicles and SDN controllers [1].

Comprehensive surveys on the software-defined networks in [41,43] lack a comparative study on the system analysis of existing SDVNs to point out SDVN components, architectures and algorithms investigated to tackle SDVN drawbacks. Authors in [41,43] investigate SDVN architectures to identify their benefits and challenges against the VANETs regarding communication in [41], and security in [43]. Within the existing SDVN solutions, technology for SDN controllers, implementation tool for the OpenFlow protocol have been proposed, yet a comprehensive study on SDVN architectures based on the modeling will provide the required insights on the components needed for further enhancements. Since the system analysis of SDVN systems provide key enabling technologies for investigating SDVENs' challenges, the solution model proposed in the implementation based the system analysis entities in SDVN shows potential research opportunities towards an efficient SDVN that could allow a huge number of next-generation VANET applications.

## 6. Conclusions

Software-defined networks (SDNs) are a network technology based on the separation of data and control planes. This paper mainly focuses on discussing implicit literature that concentrates on classifying existing SDVN solutions based on their modeling and implementation. In addition, this work provides an overview of the current studies available in the literature on C-V2X applications in support of V2X applications. The keys ideas of software-defined networks and their concepts for V2X services were also presented. We show that the simplicity of modeling that was proposed provides a detailed analysis of known solutions including SDVN or SDVN with 5G, SDVN-based cloud/mobile edge computing in order to solve current VANET issues in most cases. Loss of connectivity between vehicles and SDN controllers, routing in mobile (edge) cloud computing, were among the issues tacked by existing solutions such as SDVN, software-defined edge computing, SDVN with 5G and SDN-based routing that are currently implemented in order to solve current and ongoing VANETs challenges. Lastly, we discussed some guidelines for future research work.

**Author Contributions:** Conceptualization, writing—original draft preparation, L.N.; methodology, investigation, writing—review and editing, L.N.; formal analysis, validation, editing S.M.R.I.; supervision Y.H.C, visualization, M.B.; writing—review and editing, ressources, funding acquisition, J.J.W.

**Funding:** This research received no external funding.

**Acknowledgments:** This research was supported the BB21+ project in 2019.

**Conflicts of Interest:** The authors declare no conflict of interest.